\documentstyle[12pt,A4]{article}
\date{1-Jan-1998}
\makeatletter
\def\warnred#1#2{\@ifundefined{#1}{}{\@warning{REDEFINING \expandafter\string
\csname #1\endcsname\space TO \string#2}}}
\warnred{nc}{newcommand}\let\nc\newcommand
\nc\mycomment[1]{\marginpar{\vspace*{-1.5\baselineskip}
\tiny\begin{flushleft}#1\end{flushleft}}}
\nc\remark[1]{}
\nc\com{~,\hs}
\nc\hs{\hspace*{2em}}
\nc\mypar{\par\vskip 1ex}
\nc\noi{\noindent}
\nc\xvspace{\par~\par}
\nc\hr{\xvspace\hrule\xvspace}
\nc\ms[1]{\rule{0pt}{#1}}
\nc\SR[2]{\rule[#1]{0pt}{#2}}
\nc\vs[1]{\SR{-#1 mm}{#1 mm}}
\nc\vt[1]{\rule{0mm}{#1mm}}
\nc\vu[1]{\SR{#1 mm}{#1 mm}}
\nc\vv[1]{\SR{#1 mm}{1 mm}}
\nc\initeqn{
\def\theequation{\thesection.\arabic{equation}}\@addtoreset{equation}{section}}
\nc\sect{\mynewpage\initeqn\section}
\nc\mynewpage{{\newpage\thispagestyle{empty}\cleardoublepage
\thispagestyle{plain}}}%
\warnred{ul}{underbar}
\nc\bea{\begin{eqnarray}}
\nc\be{\begin{equation}}
\nc\bex{\begin{equation}\let\\=\com\catcode`\&=10\relax}
\nc\ee{\end{equation}}
\nc\ena{\end{eqnarray}}
\nc\bit{\begin{itemize}}
\nc\eit{\end{itemize}}
\nc\ba{\begin{array}}
\nc\ea{\end{array}}
\nc\frenchdefs[9]{
\hyphenation{ana-ly-tique ana-ly-se ana-ly-ser ca-no-nique ca-no-niques
coup-lage coup-lages de-scrip-tion diffe-rent diffe-rents diffe-rente
diffe-rentes en-cou-rage-ment en-cou-rage en-cou-rage-ments exemple
exo-tique exo-tiques fer-mio-ni-que fer-mio-ni-ques in-va-riant
in-va-riante jeune jeunes mo-di-fie mo-di-fies mo-di-fiees
in-va-riantes Karls-ruhe Le-gen-dre par-ti-cule par-ti-cules
re-nor-mali-sable re-nor-mali-sables super-espace super-champs
syme-trie syme-tries super-gravite va-ria-tion-nel va-ria-tion-nels
va-ria-tion-nelle va-ria-tion-nelles veri-fier veri-fie veri-fiant
veri-fient}
\nc\apparai[1]{appara{\^\i}#1}
\nc\cad{c'est-\`a-dire}
\nc\coder[1]{d\'e\-ri\-v\'ee#1 co\-va\-ri\-an\-te#1}
\nc\cotrans[1]{\susytrans{#1} covariante#1}
\nc\diff{diff\'e\-ren\-tiel}
\nc\diffop[1]{op\'erateur#1 \diff{#1}}
\nc\diffgeo{g\'eo\-m\'e\-trie \diff le}
\nc\eqref[1]{{\mbox{\'eq. (\protect\ref{#1})}}}
\nc\eqn{\'equation}
\nc\bewgl[1]{\eqn{#1} de mouvement}
\nc\idb[1]{identit\'e#1 de Bianchi}
\nc\MS{Mo\-d\`ele Standard}
\nc\MSS{\MS\ Su\-per\-sy\-m\'e\-tri\-que}
\nc\puisque{\hs\mbox{puisque }}
\nc\scc{super\-champ chiral}
\nc\sccs{super\-champs chiraux}
\nc\susy{super\-sym\'e\-tri}
\nc\susyalg{alg\`ebre de \susy e}
\nc\susytrans[1]{trans\-for\-ma\-tion#1 de \susy e}
\nc\vev[1]{va\-leur#1 moyenne#1 dans le vide}
\nc\wfaktor{{\kappa}}
\nc\mysect[3]{\remark{\newpage~\vfill}\sect{#1}\remark{\vfill{\Huge$$ #2 $$}\vfill}#3\remark{\newpage}}
\warnred{@oldhat}{\string\^ --- This is bad !!!}\let\@oldhat=\^
\@warning{REDEFINING \string\^ FOR \string\^i to work (@oldhat=old def)}
\def\^{\@ifnextchar i{\@oldhat\i\@gobble}{\@oldhat}}
}
\nc\savebot[1]{\addtolength\textheight{#1}}
\nc\savetop[1]{\savebot{#1}\advance\voffset-#1}
\nc\wider[1]{\advance\hoffset -#1 \advance\textwidth #1 \advance\textwidth #1}
\nc\mydraft{\def\mynewpage{\vfill\pagebreak}
	\input fullpage.sty \myheadings \savebot{2cm}}
\nc\mytwocol{\advance\columnsep 1em\twocolumn\parindent0pt}
\@ifundefined{greektex}{}{\endinput}
\newcommand\greeknewcommand[2]{\def#1{{#2}}}
\greeknewcommand{\al}{{\alpha}}
\greeknewcommand{\gm}{{\gamma}}
\greeknewcommand{\dt}{{\delta}}
\def\eps{{\epsilon}}
\greeknewcommand\veps{\varepsilon}
\greeknewcommand{\la}{{\lambda}}

\greeknewcommand\vk{\varkappa}
\def\vp{\varphi}
\greeknewcommand{\th}{\theta}
\greeknewcommand{\om}{{\omega}}
\greeknewcommand{\ze}{\zeta}

\greeknewcommand{\Gm}{\Gamma}
\greeknewcommand{\Si}{\Sigma}
\greeknewcommand{\Dt}{\Delta}
\greeknewcommand{\La}{\Lambda}
\greeknewcommand{\Th}{\Theta}
\greeknewcommand{\Om}{{\Omega}}

\greeknewcommand{\bze}{\bar\zeta}
\greeknewcommand{\thb}{\bar\theta}
\greeknewcommand{\vpb}{\bar\vp}
\greeknewcommand\bvp{\bar\varphi}%
\greeknewcommand{\lab}{\bar\lambda}
\greeknewcommand{\xib}{\bar\xi}
\greeknewcommand{\psib}{\bar\psi}
\greeknewcommand{\chib}{\bar\chi}
\greeknewcommand{\phib}{\bar\phi}
\greeknewcommand{\zeb}{\bar\zeta}

\greeknewcommand{\Phib}{\bar\Phi}
\greeknewcommand{\Psib}{\bar\Psi}
\greeknewcommand{\Thb}{\bar\Theta}
\greeknewcommand{\Pib}{\overline\Pi}
\greeknewcommand{\Lab}{\bar\Lambda}
\greeknewcommand\Sib{\bar\Si}
\greeknewcommand\Xib{\bar\Xi}

\greeknewcommand{\da}{{\dot{\alpha}}}
\greeknewcommand{\db}{{\dot{\beta}}}
\greeknewcommand{\dd}{{\dot{\delta}}}
\greeknewcommand{\dep}{{\dot{\epsilon}}}
\greeknewcommand{\de}{{\dot\varepsilon}}
\greeknewcommand{\dg}{{\dot{\gamma}}}
\greeknewcommand{\dm}{{\dot{\mu}}}

\greeknewcommand{\dv}{\dot\varphi}

\greeknewcommand{\hcq}{{\hat{\cal Q}}}
\greeknewcommand{\hB}{\hat{B}}
\greeknewcommand{\hW}{\hat{W}}
\greeknewcommand{\psih}{{\hat{\psi}}}
\greeknewcommand{\psibh}{\hs{1mm}\hat{\hs{-1mm}\psib}{}}
\greeknewcommand{\omh}{\hat{\omega}}
\greeknewcommand{\hK}{\widehat{K}}
\greeknewcommand{\omt}{\tilde{\omega}}
\greeknewcommand{\Omt}{\tilde{\Omega}}
\greeknewcommand{\hh}{\tilde{h}}
\greeknewcommand{\what}{\widehat}
\greeknewcommand{\wti}{\widetilde}
\greeknewcommand{\wGm}{{\widehat{\Gamma}}}

\greeknewcommand\undal{{\underline{\alpha}}}
\greeknewcommand\undel{{\underline{\delta}}}
\greeknewcommand\undbt{{\underline{\beta}}}
\greeknewcommand\undgm{{\underline{\gamma}}}
\greeknewcommand\unddt{{\underline{\delta}}}
\greeknewcommand\undep{{\underline{\epsilon}}}
\greeknewcommand\undsi{{\underline{\sigma}}}
\greeknewcommand\undph{{\underline{\varphi}}}
\greeknewcommand\undSi{{\underline{\Sigma}}}
\greeknewcommand\undOm{{\underline{\Omega}}}

\@ifundefined{goth}{\newfont\goth{eufm10 scaled\magstep1}}{}
\nc{\A}{{\!A}}
\nc{\bA}{{\bar{A}}}
\nc{\bD}{{\bar{D}}}
\nc{\BF}{\xb F}
\nc{\bcd}{\xb\cd}
\nc{\bch}{\overline{\cal H}}
\nc{\bfA}{{\bf A}}
\nc{\bfD}{{\bf D}}
\nc{\bff}{{\bf f}}
\nc{\bfu}{{\bf u}}
\nc{\bfv}{{\bf v}}
\nc{\bfW}{{\bf W}}
\nc{\bi}{{\bar{\imath}}}
\nc{\bj}{{\bar{\jmath}}}
\nc{\bk}{{\bar{k}}}
\nc{\bK}{{\bar{K}}}
\nc{\bl}{{\bar{l}}}
\nc{\bp}{{\bar{p}}}
\nc{\bQ}{{\bar{Q}}}
\nc{\bq}{{\bar{q}}}
\nc{\br}{{\bar{r}}}
\nc{\bR}{{\xb R}}
\nc{\bs}{{\bar{s}}}
\nc\bS{\xb S}
\nc\bT{{\hskip.1ex\overline{\hskip-.1ex T}}}%
\nc{\bU}{{\bar U}}
\warnred{bv}{bar V}
\nc\bV{\xb V}
\nc\bW{\xb W}
\nc{\bw}{{\overline{W}}}
\nc{\bcw}{{\overline{\cw}}}
\nc{\bx}{{\overline{X}}}
\nc{\bX}{{\xb{X}}}
\nc{\by}{{\overline{Y}}}
\nc{\bz}{{\bar{z}}}
\nc{\fb}{{\bar f}}
\nc{\ca}{{\cal A}}
\nc{\cb}{{\cal B}}
\warnred{cc}{\cal\space C}
\nc{\cd}{{\cal D}}
\nc{\ce}{{\cal E}}
\nc{\cf}{{\cal F}}
\nc{\cg}{{\cal G}}
\nc{\ch}{{\cal H}}
\nc{\ci}{{\cal I}}
\nc{\cj}{{\cal J}}
\warnred{ck}{\cal\space K}
\nc{\cl}{{\cal L}}
\nc{\cm}{{\cal M}}
\nc{\cn}{{\cal N}}
\nc{\co}{{\cal O}}
\nc{\cp}{{\cal P}}
\nc{\cq}{{\cal Q}}
\nc{\cR}{{\cal R}}
\nc{\cs}{{\cal S}}
\nc{\ct}{{\cal T}}
\nc{\cu}{{\cal U}}
\nc{\cv}{{\cal V}}
\nc{\cw}{{\cal W}}
\nc{\cx}{{\cal X}}
\nc{\cy}{{\cal Y}}
\nc{\cz}{{\cal Z}}
\nc{\fs}{{\bf f}}
\nc{\rd}{{\rm d}}
\nc{\re}{{\rm e}}
\nc{\rg}{{\rm g}}
\nc{\rA}{{\rm A}}
\nc{\rD}{{\rm D}}
\nc{\shalf}{\sm{1}{2}}
\nc{\undA}{{\underline{A}}}
\nc{\undB}{{\underline{B}}}
\nc{\undC}{{\underline{C}}}
\nc{\prt}{{\partial}}
\nc\LRA{\Longleftrightarrow}
\nc\impl{{\Rightarrow}}
\nc\imag{\mathop{\Im\!m}}
\nc\Lie{\mathop{\rm Lie}}
\nc\eln{\mathop{\ell\kern-.1em n}}
\nc\real{\Re\!e\,}
\nc\msh{\mbox{$\frac12$}}
\nc\mynparallel{\not{\!\|\,}}
\nc\sinc{{\rm sinc}}
\nc\tr{\mathop{\rm tr}}
\nc\Tr{\mathop{{\rm Tr}}}
\nc\Div{\partial\cdot}
\nc\dslash{\!{\not\!\partial}}
\nc\dsb{\!{\not\!\bar\partial}}
\nc\FT{{\cal F\!T}}
\nc\mybox{\square}

\nc\gf{{\gamma^5}}
\nc\unity{1\hspace{-0.25em}{\rm l}}

\nc\crc[1]{{\begin{picture}(11,9)(0,1)\put(5.5,4.5){\circle{10}}\put(2,0.8){\elvsf #1}\end{picture}}}%

\nc\bwt[1]{{\overline{\wt{#1}}}}
\nc\lra[1]{\stackrel\leftrightarrow{#1}}
\nc\vect[1]{\stackrel\rightarrow{#1}}
\nc\vecto[1]{\stackrel\longrightarrow{#1}}
\nc\wt{\widetilde}
\nc\xb[1]{{\,\overline{\!#1}}}
\nc{\sm}[2]{\hbox{\footnotesize$\displaystyle\frac{#1}{#2}$}}
\nc\sms[2]{\sm{#1}{#2}\,}
\nc{\thalf}{\tm12}
\nc{\tm}[2]{{\mbox{${#1\over#2}$}\,}}
\nc\bigexp[1]{{\mbox{\large$\rm e^{#1}$}}}
\nc\der[2]{{\partial{#1}\over\partial{#2}}}
\nc{\del}[2]{{{\delta}_{#1}}^{#2}}
\nc\diag[1]{\mathop{\rm diag}\lr(){#1}}
\nc\eqnrf[1]{\eqn{} \rf{#1}}
\nc\hstext[1]{\hs\mbox{#1}\hs}
\nc\idx[1]{\int\!{\rm d}^{#1}x\,}
\nc\inv[1]{{#1}^{-1}}
\nc\ip[1]{\iota_{#1}}
\nc\lr[3]{{\left#1 #3 \right#2}}
\nc\p[1]{{\left(#1\right)}}
\nc\mycase[1]{\left\{\begin{array}{ccl}#1\end{array}\right.}
\nc\mymat[1]{\lr(){\begin{array}#1\end{array}}}
\nc\mydbleqn[3]{\bea\begin{array}{rcl}#1
 \end{array}#2\begin{array}{rcl}#3\end{array}\ena}
\nc\mea{\@ifnextchar[{\marray}{\marray[$\displaystyle\tabskip\z@{\@@@}$&
\global\@eqcnt\@ne\hskip2\arraycolsep\hfil${\@@@}$\hfil&
\global\@eqcnt\tw@\hskip2\arraycolsep$\displaystyle\tabskip\z@{\@@@}$]}}%
\def\marray[#1]#2{\stepcounter{equation}\let\@currentlabel=\theequation
\global\@eqnswtrue\global\@eqcnt\z@\tabskip\@centering
$$\let\@@@=##
\let\\=\@eqncr\halign to \displaywidth\bgroup\@eqnsel\hskip\@centering
\tabskip\z@#1\hfil\tabskip\@centering&\llap{##}\tabskip\z@\cr
#2\@@eqncr\egroup\global\advance\c@equation\m@ne$$\global\@ignoretrue}%
\nc\set[1]{\lr\{\}{#1}}
\nc\sslash{\,/\hspace{-1.5ex}}
\nc\Slash{\,/\hspace{-1.3ex}}
\nc\SLASH[1]{\,/\hspace{-1.#1ex}}
\nc\SU[1]{{\mbox{SU$(#1)$}}}
\nc\un[1]{{\underline{#1}}}
\nc\var[2]{{\delta{#1}\over\delta{#2}}}
\nc\BBB[1]{{\Bbb #1}}
\warnred{I}{implement BBB replacement}
\def\I#1{{\rm I\!#1}}
\nc\BB{\overline B}
\nc\BG{\overline G}
\nc\CC{\mbox{\rm C\hspace{-0.55em}\sf I~}}
\nc\DD{\I D}
\nc\HH{\I H}
\nc\KK{\I K}
\nc\NN{\I N}
\nc\QQ{{\sf l\hspace{-0.4em}\rm Q}}
\nc\RR{\I R}
\nc\WW{{\sf \backslash\!\!W}}	
\nc\ZZ{{\sf Z\!\!Z}}

\nc\eu[2]{\eps^{#1 #2}\,}
\nc\sig[3]{\sigma^{#1}_{#2#3}}

\nc\dlb{{\crc B}}
\nc\dlf{{\crc F}}

\nc\0{|_{\th=0}}
\nc\ind[1]{_{\rm #1}}
\nc\dk[1]{_{(#1)}}
\nc\uk[1]{^{(#1)}}
\nc\downup[2]{_{#1}^{{\phantom{#1}}#2}}
\nc\du[2]{{}_{#1}{\kern-.1em}^{#2}}
\nc\ud[2]{{^{#1}}{\kern-.1em}_{#2}}%
\nc\dud[3]{{}_{#1}\ud{#2}{#3}}
\nc\udu[3]{{}^{#1}\du{#2}{#3}}
\nc\up[1]{{}^{#1}\!}

\nc\zerobox[2]{{\raisebox{#1}[0pt][0pt]{$\scriptstyle #2$}}}%
\nc\zerozerobox[1]{\zerobox{0pt}{#1}}%

\nc{\lsym}[1]{{\mathop{#1}\limits_{\hbox{\large$\smile$}}}}
\nc{\sym}[1]{{\mathop{#1}\limits_{\scriptstyle\smile}}}
\nc\zsym[1]{ \zerozerobox{  \mathop{#1}\limits_{\zerobox{.5ex}\smile}  } }

\nc\lsy[1]{{%
\setbox\@tempboxa\hbox{$\scriptstyle #1$}
\@tempcnta\wd\@tempboxa
\divide\@tempcnta\unitlength
\@tempcntb\@tempcnta
\divide\@tempcntb by 2\relax
\advance\@tempcnta -4\relax
\begin{picture}(0,0)
\put(\@tempcntb,-1){\oval(\@tempcnta,5)[b]}
\end{picture}
\usebox\@tempboxa
}}

\nc\gp{\up g\Phi}
\nc\gA{ \,\up g\!A}

\nc\vi[1]{\cv\ind{#1}}
\nc\li[1]{\cl\ind{#1}}
\nc\lcin{\li{cin}}
\nc\lmsq{\li{mini~SUSY~QED}}
\nc\lym{\li{YM}}
\nc\lmat{\li{mat}}
\nc\lmass{\li{mass}}
\nc\lpot{\li{pot}}
\nc\lcu{\lcin^{\rm gauge}}
\nc\lcm{\lcin^{\rm mat}}
\nc\lcl{\lcin^{\rm lin}}
\nc\lpm{\lpot^{\rm mat}}
\nc\inth[1]{\int\!{\rm d}^#1\th\,}
\nc\bpp{\lr(){\Phib\Phi}}
\nc\krs{\kappa\dk{r)(s}}
\nc\kp{K(\phi,\phib}
\nc\HHb{\overline\HH}
\nc\Phid{\Phi^\dagger}
\nc\lieco[1]{\pounds_{#1}}	
\nc\UK{{$\rm U_K(1)$}}
\nc\warg{{\{\Phi_k\}}}	
\nc\stw{{\sin\theta_W}} 
\nc\ctw{{\cos\theta_W}} 
\remark{
}
\nc\CR{\nonumber\\}
\nc\nn{\nonumber}
\nc\numbertwo{\CR[-1.5ex]\\[-1.5ex]\nn}
\nc\lbl[1]{\label{eq:#1}}
\nc\rf[1]{(\ref{eq:#1})}
\newcommand{\D}{{\cal D}}                            
\newcommand{\w}{\omega}
\newcommand{\eh}[1]{\rule{#1 cm}{0cm}}  
\newcommand{\ev}[1]{\vspace{#1 mm}}     
\newcommand{\drd}{\prt}
\newcommand{\projc}{-\frac{1}{8} \lr(){\bar\D^2 -8 R}}
\newcommand{\proja}{-\frac{1}{8} \lr(){\D^2 -8 \BR}}
\newcommand{\BA}{\overline{A}}
\newcommand{\BM}{\overline{M}}
\newcommand{\BR}{R^\dagger}

\newcommand{\BX}{\overline{X}}
\newcommand{\BW}{\overline{W}}
\newcommand{\sih}[1]{\sigma^#1}
\newcommand{\sib}[1]{\sigma_#1}
\newcommand{\sibh}[1]{{\bar \sigma}^#1}
\newcommand{\sibb}[1]{{\bar \sigma}_#1}
\newcommand{\e}[2]{e\du{#1}{#2}}        
\newcommand{\ep}[2]{e'\du{#1}{#2}}      
\newcommand{\el}{e^\lambda}             
\newcommand{\elx}[1]{e^{#1\lambda}}     

\def\j{\jmath}
\newcommand{\ph}{\phi}
\newcommand{{\bh}}{{\bar\phi}}

\newcommand{\tD}{\tilde \D}
\newcommand{\B}{{\bar{\cal D}}}                      

\def\DC{\mbox{\large$\gm$}{\kern-.1ex}}
\begin{document}
\mynewpage
%
%
%
\def\CPT#1#2{#1}
\pagenumbering{roman}
\begin{titlepage}
\makeatletter
\ev{10}
\title{\bf {Auxiliary fields rescaling in higher--derivative supergravity}}
\author{R\'egis Le D\^u\thanks{ATER \`a l'Universit\'e Pierre et Marie Curie; ~
E--mail: \tt ledu@ccr.jussieu.fr
}%
\CPT{\\
\small\em Laboratoire de Gravitation et Cosmologie Relativistes\thanks
{CNRS/Unit\'e de Recherche Associ\'ee 769},
\\
\ev{1}
\small\em Tour 22--12, 4\`eme \'etage, Bo{\^\i}te Courrier 142
\\
\ev{1}
\small\em 4,place Jussieu, 75252 PARIS Cedex 05, France. }{}%
}
\date{}
\maketitle
\vfill
\begin{abstract}
We study higher--derivative supergravity with curvature squared terms in different bases. Performing a Weyl rescaling
only on the metric or on all the superfield components does not allow to obtain a normalized kinetic Einstein term from a
$\cR+\cR^2$ theory. It is necessary to combine a Legendre transformation and a Weyl rescaling on a $\cR+\cR^2$ theory
to arrive at a theory of supergravity coupled to matter. This mechanism is applied to supergravity coupled to a
general function $k(R,\BR,\Phi,\bar\Phi)$, where $R$ is one of the supergravity chiral superfields and $\Phi$ a chiral matter
superfield.
\end{abstract}
\vfill
\phantom{Keywords: }auxiliary fields, supersymmetry breaking.\\[4ex]
{GCR--98--02--01}\\[2ex]
~\par
\vfill
\end{titlepage}
\pagenumbering{arabic}
\mynewpage


\section{Introduction}

Recently, it has been shown \cite{HOW96e} that a new mechanism for breaking supersymmetry can occur from
higher--derivative supergravity. Precisely, this mechanism is based on the equivalence between $\cR+\cR^2$
theories and gravity coupled to a scalar \cite{Cec87}. The breaking of supersymmetry clearly appears when we
analyse the scalar potential which can be done after a Weyl rescaling. The delicate point is this Weyl
rescaling that we want to discuss here. Indeed, if the Weyl rescaling is obvious in bosonic gravity, it
becomes more complicated in higher--derivative supergravity because of the auxiliary fields%
\footnote{We are interested in "old minimal" supergravity in four dimensions. Auxiliary fields are two
complex scalar fields $M, \BM$ and one real vector field $b_a$.}.
These fields, needed in the theory to close the algebra of local supersymmetry transformations, are
non--trivialy%
\footnote{The second derivative of $M$ is related to the scalar curvature $\cR$ among other terms.}
related to the metric $g_{mn}$ and the gravitino field $\psi\du{m}{\al}$. Moreover, higher--derivative
supergravity provides kinetic terms for auxiliary fields \cite{FGN78}. These terms describe new propagating degrees 
of freedom. In this context, it is important to know whether auxiliary fields have to be rescaled or not.

Our paper is organized as follows. In the second chapter we review "old minimal" supergravity. The
introduction of matter can be realized after a Weyl rescaling. Although auxiliary fields can be eliminated
by their equations of motion, it is interesting to learn what are their transformations under a Weyl
rescaling. This will be useful because it will be
not possible to eliminate auxiliary fields in higher--derivative supergravity since they appear with kinetic
terms and since the equations of motion are not linear.

The third chapter is devoted to supergravity coupled to a general function $k(R,\BR,\Phi,\bar\Phi)$ where $R$ and $\BR$ are
two covariant superfields which describe torsion and curvature \cite{WeB83}, $\Phi$ and $\bar\Phi$ are respectively
chiral and antichiral superfields. We display bosonic components of the supersymmetric lagrangian. It is shown that
the scalar curvature is not only coupled to matter but also to auxiliary fields. In this general case, we perfom a
Weyl rescaling on the metric and on all fields of the supergravity multiplet. These two rescaling do not allow for
building a linear theory of gravity without scalar curvature squared terms.

In the chapter $4$ we take place in the framework of $U(1)_K$ superspace \cite{BGG90} and consider a general 
K{\"a}hler potential $K(\phi,\bar\phi)$ where $\phi$ has a $U(1)_K$ weight $\w(\phi)$. Working in the $U(1)_K$
superspace means that one considers a superfield rescaling. It is a convenient formulation for describing matter coupled
to supergravity, but it is not well--defined when $\phi$ is replaced by $R$ with $\w(R)=2$.

Finally, in the last chapter, we explain how a $\cR+\cR^2$ theory of supergravity coupled to matter can be reduced to
a $\cR+\Pi+\Lambda$ theory of supergravity coupled to matter. First, a Legendre transformation is performed in order
that the new degrees of freedom appear explicitly. Then one can perform a Weyl rescaling on the metric, this will give
a normalized Einstein term. As an example, the case of a toy lagrangian is treated in component formulation.



\section{Rescaling in supergravity with matter}

Old minimal supergravity is described by the set of fields $(e\du ma,\psi\du m\al, M,\BM,b_a)$ where $e\du
ma$ is the vielbein field, $\psi\du m\al$ is the Rarita--Schwinger field also called gravitino, and $M,\BM,
b_a$ are auxiliary fields. An invariant lagrangian is built by taking the superspace volume \cite{WeB83}
\be
\cl_{sugra} = -3 \int d^4 \theta E~,
\label{z1}
\ee
and can be written as
\be
\cl_{sugra} = -3 \int d^2\Theta~2\ce R ~~ -3 \int d^2\bar\Theta~2\bar\ce \BR~,
\label{z2}
\ee
where $\Theta$ is a covariant variable and $\ce$ is the chiral density, with the definitions
\bea
2\ce &=& e \lr(){1+i\Theta\sih m\bar\psi_m
                  -\Theta\Theta(\BM+\bar\psi_m\bar\sigma^{mn}\bar\psi_n)}~, 
\label{z3} \\
-6R  &=& M + \Theta\lr(){\sih m\sibh n\psi_{mn}-i\sih m\bar\psi_m M +ib^m\psi_m} \CR
     & &   + \Theta\Theta \Big( -\frac 12 \cR+i\bar\psi^m\sibh n\psi_{mn}
             + \frac 23 M\BM + \frac 13 b^mb_m -i\e am\D_mb^a \CR
     & &   \eh{1}+ \frac 12 \bar\psi^2 M -\frac 12 \psi_m\sih m\bar\psi_n b^n \CR
     & &   \eh{1}+ \frac 18 \epsilon^{mnpq}\lr(){ \bar\psi_m\sibb n\psi_{pq} 
                                             + \psi_m\sib n\bar\psi_{pq} } \Big)~.
\label{z4}
\ena
$\cR$ is the scalar curvature, $\psi\du{mn}{\al}$ is defined as follows
\be
\psi\du{mn}{\al} \equiv \D_m\psi\du n\al - \D_n\psi\du m\al~,
\label{z5}
\ee
with
\be
\D_m \psi\du n\al \equiv \drd_m\psi\du n\al + \psi\du n\beta w\du{m\beta}{\al}~,
\label{z6}
\ee
where $w\du{m\beta}{\al}$ is the Lorentz connection.
Developping (\ref{z1}) in component field formulation yields
\bea
\cl_{sugra} &=&  -\frac 12  e\cR 
            +\frac 12  e\epsilon^{mnpq} \lr(){\bar\psi_m\sibb n \D_p\psi_q - \psi_m\sib n \D_p \bar\psi_q  } \CR
            & &  -\frac 13 e M\BM 
           +\frac 13 e b^a b_a~.
\label{z7}
\ena
This lagrangian describes supersymmetric Einstein gravity. It is clear that both graviton and gravitino are
propagating while $M,\BM$ and $b_a$ are not dynamical fields. The lagrangian (\ref{z1}) can be generalized by%
\footnote{In the following, the volume element $d^4 \theta$ is omitted.}
\be
\cl = \frac 12 \int {E \over R}~r ~+~ \frac 12 \int {E \over \BR}~\br~, 
\label{z8}
\ee
where $r$ is a chiral superfield with the components
\be
r|=r ~,~ \D_\al r|=\sqrt2 r_\al ~,~ \D^2 r|=-4s ~.
\label{z9}
\ee
The component field formulation of the lagrangian (\ref{z8}) is
\bea
\cl &=& \frac 12 \int {E \over R}~r ~+~ \frac 12 \int {E \over \BR}~\br \CR
    &=& \int d^2\Theta~2\ce r ~+~ \int d^2\bar\Theta~2\bar\ce \br \CR
    &=& -\frac 14 \D^2(2\ce r) -\frac 14 \bar\D^2(2\bar\ce\br) \CR
    &=& -\frac 14 \D^2(2\ce)~r -\frac 14 (2\ce) \D^2 r - \frac 12 \D^\al(2\ce)\D_\al r + h.c. \CR
    &=& -er(\BM+\bar\psi_m\bar\sigma^{mn}\bar\psi_n) + es 
        + \frac{i}{\sqrt2} e \lr(){\bar\psi_{m\da}\bar\sigma^{m\da\al}}r_\al + h.c~.
\label{z10}
\ena
The coupling of chiral matter to supergravity can be realized by considering a real function $\Omega$ of
chiral and anti--chiral superfields $\Phi$ and $\bar\Phi$
\be
\cl_{cin+mat} = -3 \int E ~ \Omega\lr(){\Phi,\bar\Phi}~,
\label{z11}
\ee
which gives in components%
\footnote{ We use (\ref{z10}) with $r=\frac{3}{8} \lr(){\B^2-8R} \Omega\lr(){\Phi,\bar\Phi}$~. }
\bea
 e^{-1}~\cl_{cin+mat} &=& ~ \Omega \lr(){-\frac 12 \cR + \frac 13 b^mb_m -\frac 13 M\BM} \CR
                      & & + \Omega_i \lr(){MF_i + ib^m\drd_m A_i} \CR
                      & & + \Omega_{\bar\imath} \lr(){\BM~\BF_i - ib^m\drd_m\BA_i} \CR
                      & & + \Omega_{i\bar\j} \lr(){-3 F_i\BF_j + 3 \drd^m A_i \drd_m \BA_j}~,
\label{z12}
\ena
using the definitions
\be
A_i         \equiv \Phi_i| ~, \eh{1} 
 \chi_{i\al} \equiv \frac{1}{\sqrt2} \D_\al \Phi_i| ~, \eh{1}
 F_i         \equiv -\frac 14 \D^2 \Phi_i| ~,
\label{z13}
\ee
and
\be
\Omega_i         \equiv {{\drd\Omega}\over{\drd A_i}}~, \eh{1}
\Omega_{\bar\imath}  \equiv {{\drd\Omega}\over{\drd \BA_i}}~, \eh{1}
\Omega_{i\bar\j} \equiv {{\drd^2\Omega}\over{\drd \BA_j\drd A_i}}~.
\label{z14}
\ee
One can show that the lagrangian (\ref{z12}) does not have a kinetic normalized Einstein term. A right
description of Einstein gravity needs the presence of $-\frac{1}{2} e\cR$ which can be obtained by performing
a rescaling on the vielbein%
\footnote{The prime denotes rescaled fields.}
\be
\e ma = \el \ep ma~,
\label{z15}
\ee
where $\lambda$ is expressed in terms of matter fields, i.e
\be
\lambda = -\frac 12 \ln \Omega~.
\label{z16}
\ee
This rescaling modifies the scalar curvature such that%
\footnote{Note that $\drd_m=\drd'_m$ and $\drd_a= e^{-\lambda} \drd'_a$.}
\be
\cR = \Omega \lr(){\cR' +6\lr(){\drd'^a\lambda C'_a+ \drd'^a\lambda\drd'_a\lambda + \drd'^a\drd'_a\lambda}}~,
\label{z17}
\ee
where
\be
C'_a \equiv \ep am \ep bn \lr(){\drd'_m\ep nb - \drd'_n\ep mb}~.
\label{z18}
\ee
Matter fields are invariant under this Weyl rescaling. This implies that the lowest components of $\Phi_i$ are
invariant
\be
A_i = A'_i~,
\label{z19}
\ee
but not highest components
\bea
 \chi\du{i}{\al} &=& e^{-\frac 12 \lambda} \chi'\du{i}{\al} ~, \CR
 F_i             &=& \elx - F'_i~.
 \label{z20}
\ena
At this stage, there are two ways to treat the auxiliary fields $M,\BM$ and $b_a$. Either one can eliminate them
by taking their equations of motion \cite{WeB83} or one can consider them as supergravity multiplet
components $(e\du ma,\psi\du m\al,M,\BM,b_a)$. In this case a superfield rescaling acts on all component
fields and one obtains  non-trivial relations between rescaled and old fields%
\footnote{We shall only keep bosonic fields.}
\bea
 M      &=& \Omega^{\frac 12} \lr(){M'+3\Omega^{-1}\Omega_{\bar\j}\BF'_j}~, \CR
 \BM    &=& \Omega^{\frac 12} \lr(){\BM'+3\Omega^{-1}\Omega_{i} F'_i}~, \CR
 b_a    &=& \Omega^{\frac 12} \lr(){b'_a -\frac{i}{2}\lr(){-3\Omega^{-1}\Omega_{\bar\imath}\drd'_a\BA_i +
 3\Omega^{-1}\Omega_i\drd'_a A_i}}~.
 \label{z21}
\ena
Then the rescaled lagrangian can be expressed in terms of new fields
\bea
 \cl_{cin+mat} &=& - \frac 12 e' \cR' 
                   + \frac 13 e' b'^m b'_m 
                   - \frac 13 e' M'\BM'  \CR
               & & + e' \lr(){-3\Omega^{-2}\Omega_i\Omega_{\bar\j} + 3\Omega^{-1}\Omega_{i\bar\j}}
              \lr(){\drd'^m A_i \drd'_m \BA_j - F'_i \BF'_j}~.
 \label{z22}
\ena
This lagrangian reproduces Einstein gravity with supersymmetric counterpart as well as a kinetic matter term.
Since $M'$ and $b_a'$ verify equations of motion
\bea
M' &=& 0~,
\CR
b_a'&=& 0~,
\label{z23}
\ena
one finds the well known lagrangian \cite{WeB83}
\be
\cl_{cin+mat} =  - \frac 12 e' \cR'
                 -e' K_{i\bar{\jmath}} \drd'^m A_i \drd'_m \BA_j
                 +e' K_{i\bar{\jmath}} F'_i \BF'_j~,
\label{z24}
\ee
with
\bea
K &=& -3 \ln \Omega~,
\CR
K_{i\bar{\jmath}} &=& \drd^2 K \over {\drd A_i \drd \bar A_{\bar{\jmath}}}~.
\label{z25}
\ena
Then one concludes that the two ways are equivalent since they give the same result. Hence, if $M, \BM$
and $b_a$ can not be replaced by their equations of motion, relations (\ref{z21}) will be useful to
rescale a non--normalized Einstein lagrangian.

In the following chapter we shall point out the difficulty of rescaling a lagrangian with
higher--derivative terms.


\section{Coupling $k(R,\BR,\Phi,\bar\Phi)$ to supergravity}

 It is well known that old minimal supergravity is completly described by a set of superfields
\cite{WeB83}:
\be
R~, \eh{.3} 
\BR~, \eh{.3}
G_a~, \eh{.3} 
W_{\lsym{\al\beta\gamma}}~, \eh{.3} 
\BW^{\lsym{\da\db\dg}}~. 
\label{z26}
\ee
In a recent paper \cite{Led97}, we analysed curvature squared terms arising from these superfields.
It appears that the highest superfield components of the combination $R\BR$ yields a $\cR^2$ term,
where $\cR$ is the scalar curvature term. In this
chapter, we generalize our study of Weyl rescaling by considering the general lagrangian
\be
\cl = -3 \int E~k\lr(){\Phi,\bar\Phi,R,\BR}~,
\label{z49}
\ee
where $k$ is a real function, $\Phi$ a chiral superfield and $R$ the superfield defined above. This
lagrangian includes the  following particular cases
\bea
\cl &=& \int E~\Omega(\Phi,\bar\Phi) f(R,\BR)~,
\label{z50}\\
\cl &=& \int E~f(R,\BR)~,
\label{z51}\\
\cl &=& \int E~e^{-\frac{K(\Phi,\bar\Phi)}{3}}~.
\label{z52}
\ena
Computing (\ref{z49}) gives
\bea
e^{-1} \cl &=&~k_{i\bar\j}~\lr\{\}{-\frac{3}{16} \D^2\phi^i \B^2\bar\phi^{\bar\j}
                                   +3 \tD^m\phi^i \tD_m\bar\phi^{\bar\j}} 
\CR
           & &+k_i~ \lr\{\}{ \frac{3}{2} R\D^2\phi^i 
                            -3i G^m\tD_m\phi^i
                            } \CR
           & &+k_{\bar\jmath}~ \lr\{\}{ \frac{3}{2} \BR\B^2\bar\phi^{\bar\j}
                            +3iG^m\tD_m\bar\phi^{\bar\j}
                            } \CR
           & &+k~~ \lr\{\}{ \frac{3}{4} \D^2 R
                            +\frac{3}{4} \B^2\BR
                            -36 R\BR}~,
\label{z53}
\ena
with the definitions
\be
\phi_i = \lr(){\Phi,R}~, \eh{1}
k_i = {\drd k\over\drd\phi^i}~, \eh{1}
k_{i\bar\j} = {\drd^2 k\over \drd\phi^i\drd\bar\phi^{\bar\j} }~,
\label{z54}
\ee
and
\bea
\tD_m \Phi &=&\drd_m \Phi~, \\
\label{z54a}
\tD_m R &=& \drd_m R +ib_m R~.
\label{z54b}
\ena
The superfield $R|=-\frac{M}{6}$  will produce the scalar curvature term, and the chiral superfields
 $\Phi$  will generate, among other terms, kinetic terms and interacting terms.

One derives the component field expression for (\ref{z53})
\bea
e^{-1} \cl &=& -\frac{1}{2} \cR 
                         \lr(){ k+Mk_M+\BM k_{\BM}-2k_{M\BM}b^ab_a-4k_{M\BM}M\BM 
                               -\frac{3}{2}\lr(){k_{\BM\Phi}F^{\Phi}+k_{M\bar\Phi}\BF^{\bar\Phi} }
                              } 
\CR
             & & - \frac{3}{4} k_{M\BM} \cR^2
                 + 3k_{M\BM} \drd^m M\drd_m \BM 
                 - 3k_{M\BM} \lr(){\e am\tD_m b^a}^2 
\CR
             & & +3 k_{M\Phi}       \drd^mM\drd_mA^\Phi
                 +3 k_{\BM\bar\Phi} \drd^m\BM\drd_m\BA^{\bar\Phi}
\CR
             & & +ib^m \lr(){k_M\drd_m M -k_{\BM}\drd_m\BM
                            +k_\Phi\drd_mA^\Phi-k_{\bar\Phi}\drd_m\BA^{\bar\Phi}
                            }
\CR
             & & -i\lr(){\e am\tD_m b^a}
                   \lr(){k_M M- k_{\BM}\BM
                        -3k_{\BM\Phi} F^\Phi+3k_{M\bar\Phi}\BF^{\bar\Phi}
                        }  
\CR
             & & -\frac{1}{3} M\BM   
                      \lr(){k-2k_M M-2k_{\BM}\BM+4k_{M\BM} M\BM
                           +6k_{\BM\Phi}F^\Phi+6k_{M\bar\Phi}\BF^{\bar\Phi}
                           } 
\CR
             & & +\frac{1}{3} b^a b_a 
                      \lr(){k+k_M M+k_{\BM}\BM-4k_{M\BM} M\BM-k_{M\BM}b^cb_c
                           -3k_{\BM\Phi}F^\Phi-3k_{M\bar\Phi}\BF^{\bar\Phi}}
\CR
             & & -3k_{\Phi\bar\Phi}F^\Phi\BF^{\bar\Phi}
                 +3k_{\Phi\bar\Phi}\drd_mA^\Phi\drd_m\BA^{\bar\Phi}
                 +k_\Phi F^\Phi M
                 +k_{\bar\Phi}\BF^{\bar\Phi} \BM~,
\label{z55}
\ena
with
\be
\tD_m b^a = \drd_m b^a + b^c w\du{mc}{a}~.
\label{z55a}
\ee
It is clear that taking $k=\Omega(\Phi,\bar\Phi)$ in (\ref{z55}) reproduces (\ref{z11}).
Starting with (\ref{z55}), a Weyl rescaling can be performed. We will consider two rescalings; either we
make the conformal transformation only on the metric
\be
\e ma = e^\lambda \ep ma~,
\label{z56}
\ee
keeping auxiliary fields inert, or we make a superconformal transformation on all fields, i.e
\bea
\e ma &=& e^\lambda \ep ma~,  \CR
M     &=& e^{-\lambda} \lr(){M'  +3k^{-1} k_{\bar\j}\BF'^{\bar\j}}~, \CR
\BM   &=& e^{-\lambda} \lr(){\BM'+3k^{-1} k_{i} F'^i}~, \CR
b_a   &=& e^{-\lambda} \lr(){b'_a-\frac{i}{2}\lr(){-3k^{-1}k_{\bar\imath}\drd'_a\BA^{\bar\imath} +
 3k^{-1}k_i\drd'_a A^i}}~,
\label{z57}
\ena
with the choice
\be
e^{2\lambda} = k^{-1}~.
\label{z58}
\ee
In the case of supergravity coupled to matter, these two conformal transformations give the same result. But
in general it is not true, since auxiliary fields can not be eliminated by equations of motion in
higher--derivative supergravity. As an example, consider the term
\be
\cl_{\cR} = \frac{1}{2} eM\BM\cR + eb^ab_a\cR~.
\label{z59}
\ee
Under transformation (\ref{z56}), this yields
\be
\cl_{\cR} = \frac{1}{2} e' e^{2\lambda} \lr(){M\BM+2b^ab_a} \lr(){\cR'+6\Delta'}~,
\label{z60}
\ee
where
\be
\Delta' = \frac{3}{4} k^{-2} \lr(){\drd'^ak\drd'_ak}
         -\frac{1}{2} k^{-1} \lr(){\drd'^a\drd'_ak}
         -\frac{1}{2} k^{-1} \lr(){\drd'^ak} \ep am \ep bn 
                             \lr(){\drd'_m\ep nb-\drd'_n\ep mb}~.
\label{z61}
\ee
The same lagrangian (\ref{z59}), under the transformation (\ref{z57}), leads to a mixing of auxiliary fields with
scalar curvature
\bea
\cl_{\cR} &=& \frac{1}{2} e' \lr(){M'  +3k^{-1}k_{\bar\jmath}\BF^{\bar\jmath}} 
                     \lr(){\BM'+3k^{-1}k_i F^i}
                     \lr(){\cR'+6\Delta'}
\CR
    & &+ e' \lr(){b'_a
                -\frac{i}{2}\lr(){-3k^{-1}k_{\bar\imath}\drd'_a\BA^{\bar\imath}}
                                  +3k^{-1}k_i\drd'_a A^i }^2 
           \lr(){\cR'+6\Delta'}~.
\label{z62}
\ena
And whatever transformation we take, (\ref{z56}) or (\ref{z57}), the  curvature squared term $\cR^2$
generates squared kinetic terms. The lagrangian
\be
\cl_{\cR^2} = -\frac{3}{4} k_{M\BM} \cR^2~,
\label{z63}
\ee
 becomes under (\ref{z56})
\be
\cl_{\cR^2} = -\frac{3}{4} k'_{M\BM} \lr(){\cR'^2+12\Delta'\cR'+36\Delta'^2}~.
\label{z64}
\ee
This lagrangian contains fourth order derivative terms in $\Phi$, which is not satisfying. Futhermore, the kinetic term
for the graviton field is not normalized, and a Weyl rescaling increases the number of couplings between the curvature
and the other fields. Moreover, this component field formulation shows us that a rescaling can not absorb $\cR^2$ terms
in the metric. Thus one has to find another way in order to rearrange higher--derivative supergravity. Another 
possibility consists in extending the symmetry group to a $U(1)$, namely $U(1)_K$ supergravity \cite{BGG90}.


\section{Superfield rescaling in $U(1)_K$ superspace}

In addition to the vielbein $E^A$ and the Lorentz connection $\phi\du BA$, the $U(1)_K$ superspace contains a $U(1)$
connection $A$ which is a one--form in superspace \cite{BGG90}
\be
A= dz^M A_M~.
\label{ww1}
\ee
$T^A$, the usual torsion of old minimal supergravity, is modified as follows
\be
T^A = dE^A + E^B \phi\du BA + \w(E^A) E^A A~,
\label{ww2}
\ee
and we define another two--form, the $U(1)$ fieldstrength $F$
\be
F=dA~.
\label{ww3}
\ee
In $U(1)_K$ supergravity, our lagrangian (\ref{z1}) becomes
\bea
\cl      &=&  -3\int E
 \CR
            &=&  -\frac 12  e\cR
                 -\frac 13  eM\BM 
                 +\frac 13  eb^a b_a
\CR 
            & &  +\frac 12  e\epsilon^{mnpq} \lr(){\bar\psi_m\sibb n \tD_p\psi_q 
                                                - \psi_m\sib n \tD_p \bar\psi_q  } 
\CR
            & &  +\frac i2 e \lr(){ \bar\psi_m\sibh m X + \psi_m\sih m \BX }|
                 -\frac 12 e \lr(){\D^\al X_\al}|~,
\label{z65}
\ena
where we define the $U(1)$ covariant derivative $\tilde \D_m X$
\be
\tilde\D_m X = \drd_m X + \w(X) \tilde A_m X~,
\label{z65b}
\ee 
with
\be
\tilde A_m = A_m + \frac{i}{2} b_m~.
\label{z65c}
\ee
Matter is included in the components of $X, \D X$ and $\tilde A_m$. Indeed, one has the relation
\bea
X_\al &=& \projc \cd_\al K(\phi,\bar\phi)~, 
\label{z66} \\
\BX^\da &=& \proja \bar\cd^\da K(\phi,\bar\phi)~.
\label{z67}
\ena
Generally, matter superfields have a $U(1)_K$ weight which is zero. One can ask%
\footnote{Our goal is to couple $R$ and $\BR$ to $U(1)_K$ supergravity. In this perspective, one has to
remember that $U(1)_K$ weight of $R$ is 2.}
what happens if we consider chiral superfields of weight $\w(\phi)$. Relation (\ref{z66}) can be written
as
\bea
X_\al &=& -\frac{1}{8} \lr(){\B^2-8R} \D_\al K
\CR
      &=& -\frac{1}{8} K_{\phi\bar\phi\bar\phi} (\B\bar\phi)^2 \D_\al\phi
\CR
      & & -\frac{1}{8} K_{\phi\bar\phi} \lr(){\B^2\bar\phi \D_\al\phi
                                              +4i\B^\da\bar\phi\tD_{\al\da}\phi}
\CR
      & & +\frac{1}{2} \w{(\phi)} K_\phi X_\al \phi~,
\label{z68}
\ena
which can be rearranged as
\be
X_\al = \lr(){1-\frac{\w(\phi)}{2} K_\phi \phi }^{-1}
        \lr(){-\frac{1}{8} K_{\phi\bar\phi\bar\phi} (\B\bar\phi)^2 \D_\al\phi
              -\frac{1}{8} K_{\phi\bar\phi} \B^2\bar\phi \D_\al\phi
              -\frac{i}{2} K_{\phi\bar\phi} \B^\da\bar\phi\tD_{\al\da}\phi
             }~.
\label{z69}
\ee
A similar relation is obtained for its complex conjugate
\be
\BX_\da = \lr(){1+\frac{\w(\bh)}{2} K_\bh \bh }^{-1}
          \lr(){-\frac{1}{8} K_{\phi\phi\bar\phi} (\D\phi)^2 \B_\da\bh
                -\frac{1}{8} K_{\phi\bar\phi} \D^2\phi \B_\da\bh
                +\frac{i}{2} K_{\phi\bar\phi} \D^\al\phi\tD_{\al\da}\bh
             }~.
\label{z70}
\ee
The lagrangian (\ref{z65}) contains a $\D^\al X_\al$ term which can be expressed as
\bea
\D^\al X_\al \lr(){1-\frac{\w(\phi)}{2} K_\phi\phi} 
&=& -\frac{1}{8} K_{\ph\ph\bh\bh} (\D\ph)^2 (\B\bh)^2
\CR
& & + K_{\ph\bh\bh} \lr(){-\frac{i}{2}\tD_{\al\da}\bh\B^\da\bh\D^\al\ph
                          -\frac{1}{8}(\B\bh)^2\D^2\ph }
\CR
& & + K_{\ph\ph\bh} \lr(){-\frac{i}{2}\tD_{\al\da}\ph\D^\al\ph\B^\da\bh
                          -\frac{1}{8}(\D\ph)^2\B^2\bh }
\CR
& & + K_{\ph\bh}    \lr(.{  -\frac{i}{2}\tD_{\al\da}\B^\da\bh\D^\al\ph }
                               +\frac{i}{2}\B^\da\bh\tD_{\al\da}\D^\al\ph
\CR
& &               \eh{1.25} -\frac{1}{8} (\D^2\ph) (\B^2\bh)
                            -\tD_{\al\da}\bh\tD^{\al\da}\ph 
\CR
& &               \eh{1.3}\lr.){  +\frac{3}{2} G_{\al\da} \B^\da\bh\D^\al\ph }
\CR
& &  +\frac 12 \lr(){\w(\ph)K_\ph 
                   + \w(\ph)K_{\ph\ph}\ph 
                   - \w(\bh)K_{\ph\bh}\bh} \lr(){\D^\al\ph X_\al}
\CR
& &  +\lr(){\w(\ph)K_{\ph\bh}\ph} \lr(){\B_\da\bh\BX^\da}~.
\label{z71}
\ena
Substituting (\ref{z69}) and (\ref{z70}) in (\ref{z71}) yields $\D^\al X_\al$ in terms of derivatives of
the chiral superfield $\phi$. Finally, $\tilde A_m$, the last component of the gauge multiplet, must be
defined. Using the definition of the fieldstrength%
\footnote{We follow the definition of \cite{WeB83}, \cite{BGG90} and \cite{Led97}.}
\be
{\tilde F}_{BA} = D_B {\tilde A}_A -(-)^{ab} D_A {\tilde A}_B -T\du{BA}{C} {\tilde A}_C~,
\label{z72}
\ee
with the constraint
\be
{\tilde F}_{\beta\da} = 0~,
\label{z73}
\ee
one deduces the expression of the vector $\tilde A_m$,
\be
\tilde A_m = \frac{1}{4} \lr(){1-\frac{\w(\ph)}{2} K_\ph\ph}^{-1}
                         \lr(){K_\ph\drd_m\ph-K_\bh\drd_m\bh
                               +\frac{i}{2} K_{\ph\bh} (\D\ph\sib m\B\bh)}~.
\label{z74}
\ee
For this last relation, we used the fact that $K_{\phi} \phi=K_{\bar\phi} \bar\phi$. Relations (\ref{z69})
to (\ref{z71}) and (\ref{z74}) allow us to express the lagrangian (\ref{z65}) in terms of supergravity
fields and derivatives of chiral field $\phi$. One finds
\bea
e^{-1}\cl &=& -\frac 12  \cR
              -\frac 13  M\BM 
              +\frac 13  b^a b_a
\CR 
          & &  +\frac 12  \epsilon^{mnpq} \lr(){\bar\psi_m\sibb n \tD_p\psi_q 
                                                - \psi_m\sib n \tD_p \bar\psi_q  } 
\CR
          & &  +\lr(){
                   \frac i2  \lambda_\phi \bar\psi_{m\da} {\bar\sigma}^{m\da\al}
                   +\Lambda^\al
                     }
               \lr(){-\frac{1}{8} K_{\phi\bar\phi\bar\phi} (\B\bar\phi)^2 \D_\al\phi
                     -\frac{1}{8} K_{\phi\bar\phi} \B^2\bar\phi \D_\al\phi
                     -\frac{i}{2} K_{\phi\bar\phi} \B^\da\bar\phi\tD_{\al\da}\phi
                    }
\CR
          & & +\lr(){
                  \frac{i}{2} \bar\lambda_\bh \psi\du{m}{\al} {\sigma}^m_{\al\da}
                  +\bar\Lambda_\da
                   }
             \lr(){-\frac{1}{8} K_{\phi\phi\bar\phi} (\D\phi)^2 \B^\da\bh
                   -\frac{1}{8} K_{\phi\bar\phi} \D^2\phi \B^\da\bh
                   +\frac{i}{2} K_{\phi\bar\phi} \D^\al\phi\tD\du{\al}{\da}\bh
                   }
\CR       & & -\frac 12  \lambda_\ph \lr{\{}{.}{ 
              -\frac{1}{8} K_{\ph\ph\bh\bh} (\D\ph)^2 (\B\bh)^2 }
\CR
& & + K_{\ph\bh\bh} \lr(){-\frac{i}{2}\tD_{\al\da}\bh\B^\da\bh\D^\al\ph
                          -\frac{1}{8}(\B\bh)^2\D^2\ph }
\CR
& & + K_{\ph\ph\bh} \lr(){-\frac{i}{2}\tD_{\al\da}\ph\D^\al\ph\B^\da\bh
                          -\frac{1}{8}(\D\ph)^2\B^2\bh }
\CR
& & + K_{\ph\bh}    \lr(.{  -\frac{i}{2}\tD_{\al\da}\B^\da\bh\D^\al\ph }
                               +\frac{i}{2}\B^\da\bh\tD_{\al\da}\D^\al\ph
\CR
& &               \eh{1.25} -\frac{1}{8} (\D^2\ph) (\B^2\bh)
                            -\tD_{\al\da}\bh\tD^{\al\da}\ph 
\CR
& &       \eh{1.3} \lr{.}{\}} {\lr.){  +\frac{3}{2} G_{\al\da} \B^\da\bh\D^\al\ph }}~.
\label{z75}
\ena
with the definitions
\bea
\lambda_\ph &\equiv& \lr(){1-\frac{\w(\phi)}{2} K_\phi \phi }^{-1}~,
\label{z76}\\
\bar\lambda_\bh &\equiv& \lr(){1+\frac{\w(\bh)}{2} K_\bh \bh }^{-1}~,
\label{z77}\\
\Lambda^\al &\equiv& \frac 12 \lambda_\ph^2
                     \lr(){\w(\ph)K_\ph 
                         + \w(\ph)K_{\ph\ph}\ph 
                         - \w(\bh)K_{\ph\bh}\bh} \D^\al\ph~, 
\label{z78}\\
\bar\Lambda_\da &\equiv& \lambda_\ph \bar\lambda_\bh
                         \lr(){\w(\ph)K_{\ph\bh}\ph} \B_\da\bh~. 
\label{z79}
\ena
Thus one obtains, in the $U(1)_K$ superspace, a lagrangian with a normalized Einstein kinetic term. Since
this result is general, it can be applied to any chiral superfield $\phi$ of $U(1)_K$ weight $\w(\phi)$.
What does (\ref{z75}) give if we decide to choose the particular case $\phi=R$ with $\w(\phi)=2$ ? In this
case, lagrangian (\ref{z75}) will contain $R|, \D_\al R|$ and $\D^2 R|$ terms with the definitions
\bea
R| &=& -\frac M6~,
\label{z46}\\
\D_\al R| &=& -\frac 16 (\sigma^n \bar\sigma^m)\du{\al}{\gamma} \lr(){\tD_n\psi_{m\gamma} - \tD_m\psi_{n\gamma}} \CR
          & & -\frac i6 b^m \psi_{m\al} 
              +\frac i6 \lr(){\sigma^m \bar\psi_m}_\al M
              -\frac 13 X_\al~,
\label{z47}\\
\D^2 R| &=& -\frac 13 \cR
            +\frac 29 b^ab_a
            +\frac 49 M\BM
            -\frac{2i}{3} \e am\tD_m b^a \CR
        & & +\frac{2i}{3} \lr(){\tilde\psi_{mn}\sigma^m\bar\psi^n}
            -\frac{1}{3}  \lr(){\psi_m\sigma^m\bar\psi^n}b_n
            +\frac{1}{3}  \lr(){\bar\psi^m\bar\psi_m}M \CR
        & & +\frac{1}{12} \eps^{mnpq} \lr(){\tilde\psi_{mn} \sigma_n\bar\psi_q + \psi_m\sigma_n\tilde{\bar\psi}_{pq} }
            -\frac{1}{3} \D^\al X_\al \CR
        & & +\frac{i}{3} \lr(){X\sigma^m\bar\psi_m - \BX\bar\sigma^m\psi_m}~.
\label{z48}
\ena
It is clear that the two last terms will generate $\D^\al X_\al$, $\BX_\da$ and $X_\al$ terms in the lagrangian which 
we do not want.

The problem comes from the expression of $X_\al$ (\ref{z69}), $\BX_\da$ (\ref{z70}) and $\D^\al X_\al$
(\ref{z72}) which are not defined for $\phi=R$. Indeed, one has
\bea
X_\al = & &\lr(){1-K_RR}^{-1} \times \CR
        & &\lr(){-\frac{1}{8} K_{R\BR\BR} (\B\BR)^2 \D_\al R
                 -\frac{1}{8} K_{R\BR} \B^2\BR \D_\al R  
                 -\frac{i}{2} K_{R\BR} \B^\da\BR \tD_{\al\da} R
             }~.\CR
        & &
\label{z80}
\ena
In this expression, the right hand side involves terms such as $\BX^2, X_\al$ and $\D^\al X_\al$. Thus, one can
not explicitly express quantities $X_\al, \BX_\da$ and $\D^\al X_\al$ as well as $\tilde A_m$ in terms of
independent fields. Conformal transformation is not possible in $U(1)_K$ superspace with a function
$K(R,\BR)$ because $R$ and $\BR$ superfields can not be treated at the same time as matter fields and as
fields belonging to the supergravity multiplet.


\section{Legendre transformation and superfield redefinition}

Our purpose is to obtain a normalized Einstein term from higher--derivative supergravity coupled to matter.
Since a Weyl rescaling and a superfield rescaling are not allowed in higher--derivative supergravity, another approach
has to be found. A correct description of such theories can be realized by a combination of two
transformations: a Legendre transformation and a superfield redefinition. This was already done in
\cite{HOW96f}. We generalize to the case of supergravity coupled to a general function $k(\Phi,\bar\Phi,R,\BR)$. 
We start from a $\cR+\cR^2$ theory of supergravity%
\footnote{This lagrangian is displayed in (\ref{z55}).}
\be
\cl_1 = -3 \int E~k(\Phi,\bar\Phi,R,\BR)~,
\label{x1}
\ee
which is equivalent to a theory of supergravity coupled to two more matter superfields $\Pi$ and $\Lambda$,
\be
\cl_2 = -3 \int E~k(\Phi,\bar\Phi,\Pi,\bar\Pi)
        +\lr[]{-3\int {E\over R} \Lambda \lr(){R-\Pi}+ h.c.}~,
\label{x2}
\ee
in the sense that superfield equations of motion give
\be
\Pi = R~.
\label{x3}
\ee
This second theory has a non--normalized Einstein term and can be reduced to a third theory%
\footnote{We take $\phi_i=\lr(){\Phi,\Pi,\Lambda}$ and $A_i=\phi_i|$, $F_i=-\frac{1}{4} \D^2\phi_i$.}
\be
\cl_3 = -\frac{1}{2} e'\cR'
        -e' K_{i\bar{\jmath}} \drd'^m A_i \drd'_m \BA_j
        +e' K_{i\bar{\jmath}} F'_i \BF'_j~,
\label{x4}
\ee
with a field redefinition
\be
\Phi'=\Phi~, \eh{1} \Pi'=\Pi~, \eh{1} \Lambda'=\Lambda~,
\label{x5}
\ee
and
\be
\ep ma = e^{-\lambda} \e ma~,
\label{x6}
\ee
with
\be
\lambda = -\frac 12 \ln k~.
\label{x7}
\ee
At this stage we would like to make an important remark. In the previous section we tried to rescale higher--derivative
supergravity with  scalar curvature squared term. We have seen two problems. The first one was the presence of a $\cR^2$
term  which could not be absorbed. A Legendre transformation solves this problem by introducing new fields. The
other problem was the rescaling of a $K(R,\BR)$ function. In this formulation there is no such question, since  the
superfield $R$ is hidden in $\Pi$, and $\Pi$ is treated as a matter superfield. Thus one can say that the transformation
$\cl_2 \longrightarrow \cl_3$ is not a superWeyl rescaling but a field redefinition where some of supergravity fields
are rescaled $(e\du ma,\psi\du m\al,M,\BM,b_a)$, but a certain combination of them is invariant because
$\Pi=R=-\frac{M}{6}$ is considered as matter field and is a invariant superfield ($\Pi\longrightarrow\Pi$ under this
redefinition).
\ev{5}\\
As an example, we propose to construct a particular function $f(R,\BR)$ coupled to supergravity. We consider the case 
\cite{HOW96e}
\be
f(R,\BR) = f(R) + f(\BR)~,
\label{z27}
\ee
An invariant lagrangian is
\be
\cl_1 = -3 \int E f(R,\BR)~.
\label{z28}
\ee
Taking $r=-6Rf(R)$ in the generic construction (\ref{z10}), one derives the component field
expression
\bea
e^{-1} \cl_1 &=& -\frac 12 \cR      \lr(){ f + M f_M +\BM f_{\BM} }
                 +\frac 13 b^a b_a  \lr(){ f + M f_M +\BM f_{\BM} } \CR
             & & -\frac 13 M\BM     \lr(){ f -2M f_M -2\BM f_{\BM} }
                 +ib^m              \lr(){ f_M \drd_m M - f_{\BM} \drd_m \BM }\CR
             & & +i\e am \D_m b^a   \lr(){ f_{\BM} \BM - f_M M }~.
\label{z29}
\ena
In this lagrangian the scalar curvature is coupled to auxiliary fields $M$ and $\BM$.
Moreover, derivative terms of $M,\BM$ and $b_a$ are present, but we can not consider them as
propagating fields for instance. One can treat the $b_a$ field as a purely auxiliary field, it
means that $b_a$ can be replaced by its equations of motion. We start by integrate $\tD_m b_a$ term.
One has the relation
\be
e\e am \D_m v^a = \drd_m(e v^a \e am) + \frac{ie}{2} v^a (\e bn\e am-\e bm\e an)(\psi_n\sih b\bar\psi_m)~,
\label{z30}
\ee
where we choose
\be
v^a=b^a(f_{\BM} \BM-f_M M)~.
\label{z31}
\ee
Since we are only interested in bosonic terms, we simply have
\bea
e\e am\D_m \lr(){b^a(f_{\BM} \BM-f_M M)} &=& e\e am\D_m b^a~(f_{\BM} \BM-f_M M) \CR
                                         & & + e\e am b^a~(f_{\BM}\drd_m\BM + f_{\BM\BM}\BM\drd_m\BM 
                                                         -f_M\drd_m M - f_{MM} M\drd_m M) \CR
                                         &=& \mbox{total derivative}~,
\label{z32}
\ena
i.e~:
\be
e\e am\D_m  b^a \lr(){f_{\BM} \BM-f_M M} = -e\e am b^a \lr(){ f_{\BM}\drd_m\BM + f_{\BM\BM}\BM\drd_m\BM
                                                             -f_M\drd_m M - f_{MM} M\drd_m M}~.
\label{z33}
\ee
Substituting (\ref{z33}) in (\ref{z29}) gives
\bea
e^{-1} \cl_1 &=& -\frac 12 \cR      \lr(){ f + M f_M +\BM f_{\BM} }
                 +\frac 13 b^a b_a  \lr(){ f + M f_M +\BM f_{\BM} } \CR
             & & -\frac 13 M\BM     \lr(){ f -2M f_M -2\BM f_{\BM} }
                 +2ib^m             \lr(){ f_M \drd_m M - f_{\BM} \drd_m \BM }\CR
             & & +i b^m             \lr(){ f_{MM} M\drd_m M - f_{\BM\BM} \BM\drd_m\BM}~.
\label{z34}
\ena
As we consider $b_a$ as an auxiliary field, one can diagonalize contributions of this field. One
defines
\be
{\hat b_m} \equiv b_m +\frac{3i}{2} \lr(){f+Mf_M+\BM f_{\BM}}^{-1} 
                                    \lr(){(f_{MM} M+2f_M)\drd_m M - (f_{\BM\BM} \BM+2f_{\BM})\drd_m\BM}~,
\label{z35}
\ee
and the lagrangian becomes
\bea
e^{-1} \cl_1 &=& -\frac 12 h\cR 
                 -\frac 34 h^{-1} \lr(){(f_{MM} M+2f_M)\drd_m M - (f_{\BM\BM} \BM+2f_{\BM})\drd_m\BM}^2 \CR
             & & -\frac 13 M\BM   \lr(){f-2Mf_M-2\BM f_{\BM}} 
                 +\frac 13 h {\hat b^a}{\hat b_a} \CR
             & & +\frac 32 h^{-1} \lr(){(f_{MM} M+2f_M)\drd_m M - (f_{\BM\BM} \BM+2f_{\BM})\drd_m\BM}^2~,
\label{z36}
\ena
where
\be
h\equiv f+Mf_M + \BM f_{\BM}~.
\label{z37}
\ee
In order to study the breaking of supersymmetry, one has to compute the scalar potential. This can
be done after a Weyl rescaling, because this lagrangian clearly exhibits a non normalized Einstein
term. The conformal transformation
\be
\e ma \equiv \el \ep ma~,
\label{z38}
\ee
with
\be
\elx 2 \equiv h^{-1} \equiv \lr(){f+Bf_B + \bar B f_{\bar B}}^{-1}~,
\label{z39}
\ee
where we take $B=M$,
gives a correct Einstein term $-\frac{1}{2} e'\cR'$. Following the method of \cite{HOW96e}, i.e
performing a Weyl rescaling only on the metric and not on auxiliary fields, one obtains
\bea
e'^{-1} \cl_3 &=& -\frac 12 \cR' 
                  -3 h^{-2} \lr(){(f_{BB} B+2f_B)\drd'_m B (f_{\bar B\bar B} \bar B+2f_{\bar B})\drd'^m\bar B} \CR
              & & +\frac 13 h^{-1} {\hat b^a}{\hat b_a} 
                  -\frac 13 B\bar B h^{-2} \lr(){f-2Bf_B-2\bar B f_{\bar B}}~.
\label{z40}
\ena
As a remark we can say that this transformation is a field redefinition with $B=M$ and $\ep ma \equiv e^{-\lambda}\e ma$.
 In this sense it is not a Weyl rescaling but a field redefinition.\\

In other words, this lagrangian can be written as
\be
\cl_3 = -\frac 12 e' \cR' -e' K_{B\bar B} \drd'^m B\drd'_m\bar B -e'\cv(B,\bar B)+e'\frac 13 h^{-1} {\hat b^a}{\hat b_a} ~,
\label{z41}
\ee
with the following definitions
\bea
K(B,\bar B) &=& -3\ln h~,
\label{z42}\\
K_B &=& -3 h^{-1} h_B~,
\label{z43}\\
K_{B\bar B} &=& 3 h^{-2} h_B h_{\bar B} -3h^{-1} h_{B\bar B}~.
\label{z44}
\ena
$\cv(B,\bar B)$ is the scalar potential and is given by
\be
\cv(B,\bar B) = \frac 13 h^{-2} B\bar B \lr(){f-2Bf_B -2\bar B f_{\bar B}}~.
\label{z45}
\ee
 Thus one deduces that lagrangian (\ref{z28}) is equivalent to
supergravity coupled to a complex scalar field, namely $B$.

At the superfield level and in this particular case, where $f(R,\BR)=f(R)+f(\BR)$, one can say that this theory is 
equivalent to supergravity coupled to only one chiral superfield $\Pi$. Indeed, the superfield formulation of this
example is more simple than (\ref{x2}). On the one hand, the chiral matter superfields $\Phi$ are absent, and on the
other hand the $\Lambda$ superfield can be expressed in terms of $\Pi$ and $f(\Pi)$ in this special case \cite{HOW96e}.
This reduction to one superfield instead of two chiral superfields $\Pi$ and $\Lambda$, can be understood in component
formulation: one of the auxiliary fields, $b_a$, can be eliminated by its equations of motion (\ref{z33}). This is the
case when the lagrangian does not contain any fourth order derivative such as $(\D b_a)^2$ or $(\drd M)^2$.

In general both auxiliary fields, $b_a$ and $M$, can not be replaced by their equations of motion in higher--derivative
supergravity. This is why two chiral superfields are needed to construct a theory of supergravity with a normalized
kinetic Einstein term.


\section{Conclusion}

We emphasized the fact that different bases can be used in higher--derivative supergravity with  scalar curvature squared
term. An appropriate formulation consists in performing not only a Weyl rescaling, but also a Legendre transformation. We
built a general theory of higher--derivative supergravity coupled to matter and displayed the bosonic component
lagrangian. This theory is equivalent to Einstein supergravity coupled to matter with two additional chiral superfields.
It is better to work in this formulation since the scalar potential can be easily calculated in order to see whether
supersymmetry is spontaneously broken or not.

We have shown that the auxiliary fields ($M,\BM,b_a$) are not invariant under a Weyl rescaling. When we work at the superfield level
the whole set of fields $(e\du ma,\psi\du m\al,M,\BM,b_a)$ has to be modified if a Weyl rescaling is performed. Futhermore,
we studied $U(1)_K$ supergravity where we have constructed a general lagrangian with a chiral
superfield $\phi$ of weight $\w(\phi)$. As it well known, the formulation of supergravity in $U(1)_K$
superspace presents a natural framework for describing matter coupled to supergravity \cite{BGG90}. In this
formulation $R$ has a weight two, and, unfortunately, when we replace this field in our general calculation, one can see
that the fields $X, \D X$ and $\tilde A_m$ can not be expressed in terms of the supergravity multiplet 
$(e\du ma,\psi\du m\al,M,\BM,b_a)$. 

We expect that higher--derivative supergravity will help us to understand the role played by supergravity 
auxiliary fields in supersymmetry breaking.

%

\section*{Acknowledgements}

I would like to thank P.Bin\'etruy for suggestions and discussions about this work.

\mynewpage


\def\bibname{References}
\addcontentsline{toc}{section}{\bibname}



\end{document}